\documentclass[letterpaper, 10 pt, conference]{ieeeconf}  

\IEEEoverridecommandlockouts                              

\usepackage{graphics} 
\usepackage{epsfig} 
\usepackage{mathrsfs}
\usepackage{amsmath} 
\usepackage{amssymb}  
\usepackage{float}    
\usepackage[hidelinks]{hyperref}
\usepackage{algorithm}
\usepackage{booktabs}
\usepackage{xcolor}
\usepackage{algpseudocode} 

\algrenewcommand\alglinenumber[1]{\footnotesize [#1]}

\title{\LARGE \bf
DeePC vs. Koopman MPC for Pasteurization: A Comparative Study
}

\author{Branislav Dar\'a\v{s}$^{1}$ -- Patrik Val\'abek$^{2}$ -- Martin Klau\v{c}o$^{1}$
\thanks{$^{1}$Authors are with the Department of Control Engineering, Faculty of Electrical Engineering, Czech Technical University in Prague, Karlovo N\'am\v{e}st\'i 13, 120 00 Prague 2, Czech Republic
        {\tt\small \ttfamily\{branislav.daras, martin.klauco\}@cvut.cz}}%
\thanks{$^{2}$Author is with the Slovak University of Technology in Bratislava, Institute of Information Engineering, Automation, and Mathematics, Radlinsk\'eho 9, Bratislava, Slovakia
        {\tt\small \ttfamily patrik.valabek@stuba.sk}}}%

\begin{document}

\maketitle
\thispagestyle{empty}
\pagestyle{empty}

\begin{abstract}

Data-driven predictive control methods can provide the constraint handling and optimization of model predictive control (MPC) without first-principles models. Two such methods differ in how they replace the model: Data-enabled predictive control (DeePC) uses behavioral systems theory to predict directly from input--output trajectories via Hankel matrices, while Koopman-based MPC (KMPC) learns a lifted linear state-space representation from data. Both methods are well studied on their own, but head-to-head comparisons on multivariable process control problems are few. This paper compares them on a pasteurization unit with three manipulated inputs and three measured outputs, using a neural-network-based digital twin as the plant simulator. Both controllers share identical prediction horizons, cost weights, and constraints, so that differences in closed-loop behavior reflect the choice of predictive representation. Results show that both methods achieve feasible constrained control with comparable tracking error, but with a clear trade-off: KMPC tracks more tightly under the chosen cost, while DeePC produces substantially smoother input trajectories. These results help practitioners choose between the two approaches for thermal processing applications.

\end{abstract}

\section{INTRODUCTION}

Pasteurization is a safety- and quality-critical thermal process that must track setpoints reliably while respecting actuator limits and output bounds \cite{ref:pasteurization_process}. In practice, pasteurization units are multivariable and strongly coupled: changes in heating, cooling, and flow-related actuators affect multiple temperatures and quality-relevant variables simultaneously \cite{ref:thermal_systems}. Building and maintaining a high-fidelity first-principles model for such units is often time-consuming and error-prone due to unmodeled heat losses, varying operating conditions, and plant-specific effects \cite{ref:process_modeling}.

Model predictive control (MPC) is a natural candidate for such applications, as it explicitly optimizes tracking and control effort over a prediction horizon while enforcing hard constraints \cite{ref:mpc_textbook,ref:mpc_survey,ref:mpc_recent}. However, MPC performance depends on an adequate predictive model and often on state estimation \cite{ref:mpc_state_estimation}. For a coupled thermal unit, obtaining such a model requires substantial effort---equation derivation, parameter identification, validation, and re-tuning as conditions change. These practical issues motivate data-driven methods that preserve the benefits of MPC while reducing dependence on first-principles modeling \cite{ref:data_driven_control_survey}.

Two data-driven predictive control approaches have gained traction. Both bypass first-principles modeling, but they differ in what they do with data:
\begin{itemize}
	\item \textbf{DeePC} \cite{coulson_deepc_arxiv}: data $\to$ Hankel matrices $\to$ constrained optimization. No model is constructed. No state is defined. No observer is needed. The raw input--output trajectories, organized via behavioral systems theory \cite{ref:behavioral_systems} and Willems' fundamental lemma \cite{ref:willems_lemma}, are the predictor.
	\item \textbf{KMPC} \cite{ref:koopman_modeling,ref:koopman_deepc}: data $\to$ neural network training $\to$ Koopman matrices $(A_{\mathcal{K}}, B_{\mathcal{K}}, C_{\mathcal{K}})$ $\to$ lifted state-space model $\to$ Kalman filter $\to$ standard MPC. This is a full identify-then-control pipeline---the model is learned from data rather than derived from physics, but the control design follows the classical model-based path.
\end{itemize}
This is not a difference in tuning or implementation. It is a difference in architecture: DeePC eliminates the model step; KMPC automates it. Both approaches have been studied individually---DeePC with extensions to robust and regularized settings \cite{ref:berberich_stability,ref:depersis_formulas,ref:select_dpc}, and KMPC with offset-free formulations \cite{ref:offset_free_mpc,ref:koopman_deepc}---but direct comparisons on multivariable process control benchmarks are few.

We compare DeePC and KMPC on a multivariable pasteurization unit with three inputs and three outputs, simulated via a neural-network-based digital twin. Both controllers share identical prediction horizons, cost weights, and constraint sets, so that differences in closed-loop behavior reflect the predictive representation, not the tuning. The contributions are:
\begin{itemize}
	\item A formulation of regularized DeePC with practical design choices for Hankel construction, regularization, and constraint handling, applicable to multivariable systems under standard assumptions.
	\item A comparison of DeePC and KMPC on a coupled thermal process, evaluated through tracking accuracy, control effort, constraint satisfaction, and energy consumption.
	\item A documented trade-off: KMPC tracks more tightly under the chosen cost, while DeePC produces substantially smoother actuation. Both maintain comparable tracking error and constraint feasibility.
\end{itemize}

The remainder of the paper is organized as follows. Section~\ref{sec:problem_statement} describes the pasteurization plant and states control objectives. Section~\ref{sec:deepc_section} presents the DeePC formulation and the Koopman baseline. Section~\ref{sec:sim_experiments} details the simulation setup, metrics, and comparative results. The paper concludes with observations on choosing between these data-driven strategies for thermal processing.

\section{Problem Statement and Control Objectives}
\label{sec:problem_statement}

\subsection{Problem Formulation}

Consider a discrete-time system with $n_{\mathrm{u}}$ inputs and $n_{\mathrm{y}}$ outputs,
\begin{equation}
	y_k = f(y_{k-1}, \ldots, y_{k-n}, u_{k-1}, \ldots, u_{k-n}),
	\label{eq:system_io}
\end{equation}
where $u_k \in \mathbb{R}^{n_{\mathrm{u}}}$ and $y_k \in \mathbb{R}^{n_{\mathrm{y}}}$. The control objective is to track a reference signal $r_k \in \mathbb{R}^{n_{\mathrm{y}}}$ while satisfying pointwise constraints $u_k \in \mathcal{U}$ and $y_k \in \mathcal{Y}$ at every time step, and minimizing a combined cost of tracking error and control effort. In addition, we evaluate energy consumption as a secondary criterion, since reducing energy use is a direct concern in industrial pasteurization.

We do not assume that a parametric model of \eqref{eq:system_io} is available. Instead, both controllers compared in this paper are designed from input--output data alone. The following assumptions are made:
\begin{enumerate}
	\item The underlying system is linear time-invariant (LTI), or operates in a regime where the LTI approximation is adequate.
	\item The system is controllable and observable.
	\item A dataset $(u^{\mathrm{d}}, y^{\mathrm{d}})$ of sufficient length and richness is available, satisfying the persistent excitation condition required by Willems' lemma (stated in Section~\ref{sec:deepc_willems}).
	\item The reference trajectory is feasible, i.e., the system can reach and be stabilized in a neighborhood of the reference under the given constraints.
\end{enumerate}
Under these assumptions, both DeePC and KMPC can be formulated as receding-horizon constrained optimization problems, differing only in how the predictive model is constructed from data. The precise formulations are given in Section~\ref{sec:deepc_section}.

\subsection{Case Study Description}

\begin{figure}[t]
  \centering
  \includegraphics[width=\columnwidth]{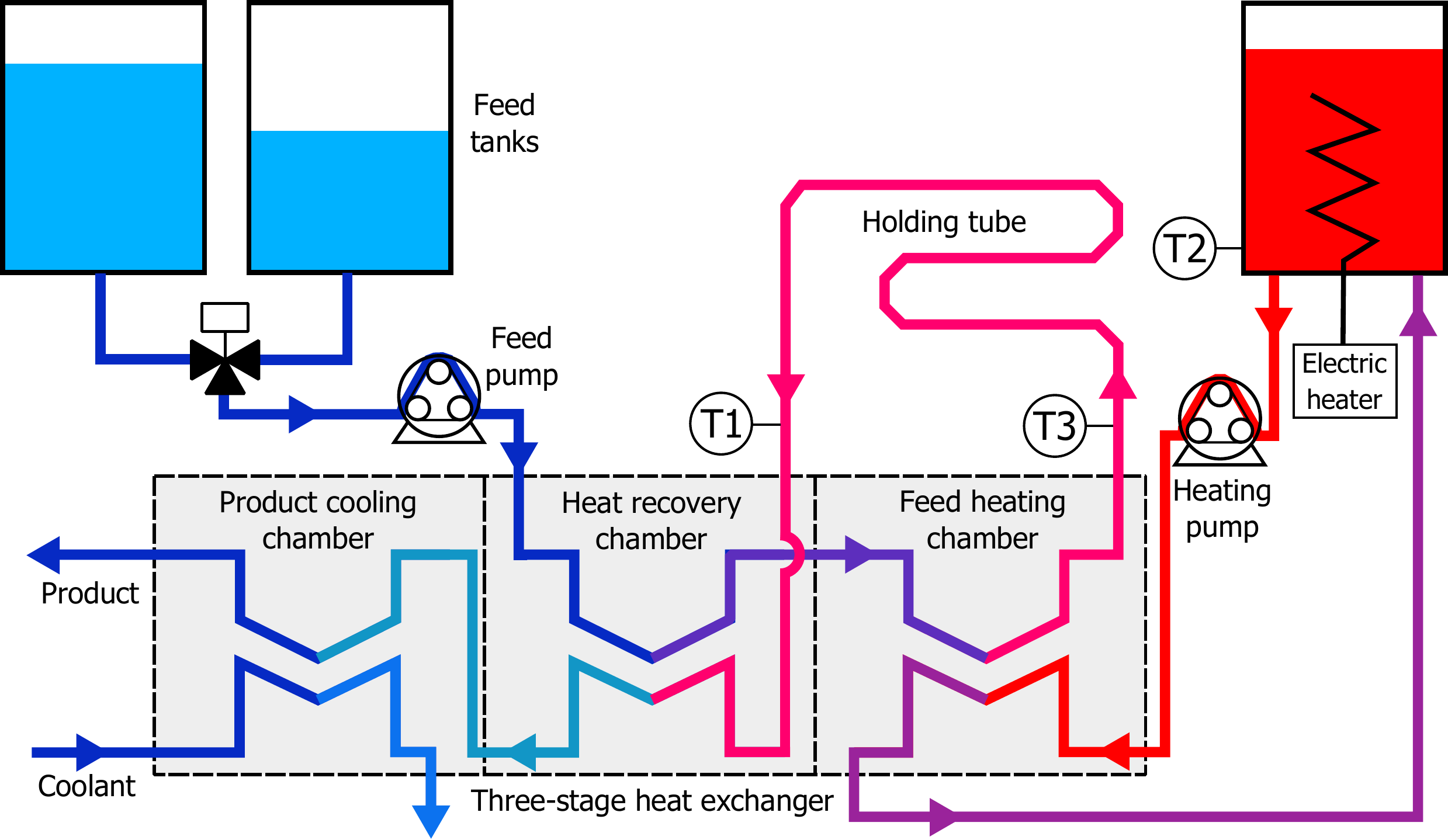}
  \caption{Pasteurization unit scheme. Manipulated inputs: \(u_1\) feed (product) flow rate, \(u_2\) hot-water circulation flow rate, \(u_3\) electric-heater power. Measured outputs: \(y_1\) holding-tube outlet temperature \(T_1\), \(y_2\) hot-tank temperature \(T_2\), \(y_3\) heat-exchanger outlet temperature \(T_3\).}
  \label{fig:pu-scheme}
\end{figure}

The case study considers a laboratory-scale continuous pasteurization unit (Fig.~\ref{fig:pu-scheme}) built around a plate heat exchanger with heat-recovery, heating, and cooling sections. Cold product is driven by a feed pump through the recovery stage, where it is preheated by the outgoing pasteurized stream, and then through the main heating stage supplied by a recirculating hot-water loop. The hot-water temperature is sustained by an electrically heated tank. 
After the product exits the heat exchanger (temperature $T_3$), it passes through an insulated holding tube to provide the required thermal residence time before returning through recovery and final cooling. As for the controlled variable, we are interested in the holding-tube outlet temperature $T_1$, which is the most important for safety and quality control.
From the extended experimental data, the neural network-based digital twin model was constructed similarly to~\cite{ref:valabek2025deep}. The model is used to simulate the plant behavior for the control design and evaluation.

\section{Data-Enabled Predictive Control}
\label{sec:deepc_section}
This section focuses on the DeePC formulation used in this study. Case-study motivation is provided in Section~\ref{sec:problem_statement}, and benchmark setup is detailed in Section~\ref{sec:sim_experiments}. We first introduce DeePC and then summarize the Koopman baseline used for comparison.

\subsection{DeePC}
\label{sec:deepc}
DeePC is an optimization-based data-driven control approach that computes constrained control actions directly from input--output trajectories, without constructing a state-space model \cite{coulson_deepc_arxiv}. We present the behavioral-systems foundation (Willems' lemma), the Hankel matrix construction, and the regularized DeePC optimization problem.

\subsubsection{Behavioral Systems Theory and Willems' Lemma}
\label{sec:deepc_willems}
Instead of a parametric state-space model, DeePC builds a non-parametric predictor from measured input--output trajectories \cite{coulson_deepc_arxiv}. Its theoretical foundation is behavioral systems theory \cite{ref:behavioral_systems}, which characterizes dynamical systems by their external behavior---the set of input--output trajectories---rather than state-space equations. For a discrete-time LTI system with $n_{\mathrm{u}}$ inputs and $n_{\mathrm{y}}$ outputs, the behavior $\mathcal{B}$ is the set of all trajectories $(u,y)$ consistent with the system dynamics.

Willems' fundamental lemma \cite{ref:willems_lemma} provides the key data-driven result. If a sufficiently long and persistently exciting trajectory $(u^{\mathrm{d}}, y^{\mathrm{d}})$ is available, any trajectory in $\mathcal{B}$ can be represented as a linear combination of Hankel-matrix columns. Persistent excitation requires sufficiently rich input signals so that the Hankel matrix has full row rank. Let $T_{\mathrm{ini}}$ denote the length of the past window used for initialization and $N$ the prediction horizon. The minimum required data length is $T \geq (n_{\mathrm{u}}+1)(T_{\mathrm{ini}}+N+n(\mathcal{B}))-1$, where $n(\mathcal{B})$ is the lag of the behavior (system order), typically unknown in data-driven settings. The input must excite at least $n(\mathcal{B})+T_{\mathrm{ini}}+N$ modes \cite{ref:willems_lemma}.

\subsubsection{Hankel Matrix Construction}
Given a dataset of length $T$: $\{u^{\mathrm{d}}_{0},\ldots,u^{\mathrm{d}}_{T-1}\}$, $\{y^{\mathrm{d}}_{0},\ldots,y^{\mathrm{d}}_{T-1}\}$ with $u^{\mathrm{d}}_{k} \in \mathbb{R}^{n_{\mathrm{u}}}$ and $y^{\mathrm{d}}_{k} \in \mathbb{R}^{n_{\mathrm{y}}}$, define the Hankel matrix \cite{ref:hankel_matrices}:
\begin{equation}
\mathscr{H}_{L}(u^{\mathrm{d}}) =
\begin{bmatrix}
u^{\mathrm{d}}_{0} & u^{\mathrm{d}}_{1} & \cdots & u^{\mathrm{d}}_{T-L} \\
u^{\mathrm{d}}_{1} & u^{\mathrm{d}}_{2} & \cdots & u^{\mathrm{d}}_{T-L+1} \\
\vdots & \vdots & \ddots & \vdots \\
u^{\mathrm{d}}_{L-1} & u^{\mathrm{d}}_{L} & \cdots & u^{\mathrm{d}}_{T-1}
\end{bmatrix}.
\label{eq:hankel_def}
\end{equation}
We partition the Hankel matrix into past ($T_{\mathrm{ini}}$ rows) and future ($N$ rows) blocks: $\mathscr{H}_{L}(u^{\mathrm{d}}) = [U_{\mathrm{p}}^\top\ U_{\mathrm{f}}^\top]^\top$, and similarly for $y^{\mathrm{d}}$. Here, $L = T_{\mathrm{ini}} + N$ is the total Hankel depth. The coefficient vector $g \in \mathbb{R}^{T-L+1}$ selects a trajectory from data \cite{coulson_deepc_arxiv}. The past blocks $U_{\mathrm{p}} \in \mathbb{R}^{(n_{\mathrm{u}} T_{\mathrm{ini}}) \times (T-L+1)}$ and $Y_{\mathrm{p}} \in \mathbb{R}^{(n_{\mathrm{y}} T_{\mathrm{ini}}) \times (T-L+1)}$ encode initial conditions. The future blocks $U_{\mathrm{f}} \in \mathbb{R}^{(n_{\mathrm{u}} N) \times (T-L+1)}$ and $Y_{\mathrm{f}} \in \mathbb{R}^{(n_{\mathrm{y}} N) \times (T-L+1)}$ represent predictions over the control horizon.

\subsubsection{Regularized DeePC}
\label{sec:deepc_regularized}
At time $t$, let $(u_{\mathrm{ini}},y_{\mathrm{ini}})$ denote the most recent past window of length $T_{\mathrm{ini}}$. In the ideal, noise-free case with exact satisfaction of Willems' lemma conditions, DeePC finds trajectories $(u,y)$ that minimize tracking error and control effort while exactly matching past observations and lying in $\mathcal{B}$ \cite{coulson_deepc_arxiv}. In practice, measurement noise can make exact past-output matching infeasible, and the Hankel-based equality constraint can be ill-conditioned when $T-L+1$ is large. To handle both issues, we use regularized DeePC, which adds a slack variable $\sigma_y \in \mathbb{R}^{n_{\mathrm{y}}T_{\mathrm{ini}}}$ for soft past-output matching and ridge regularization on $g$ for numerical stability \cite{coulson_deepc_arxiv,ref:regularization}.

The regularized DeePC optimization problem is:
\begin{subequations}
	\label{eq:deepc_reg}
	\begin{eqnarray}
		\label{eq:deepc_reg_objective}
                \hspace{-2.0em}&\min\limits_{g,\,u,\,y,\,\sigma_{y}}&\sum_{k=0}^{N-1} \left(\|y_k-r_{t+k}\|_{Q}^{2} + \|\Delta u_k\|_R^{2}\right) \\
                \hspace{-2.0em}&~&+\ \lambda_{g}\,\|g\|_{2}^{2} + \lambda_{\sigma}\,\|\sigma_{y}\|_{2}^{2}, \nonumber \\
		\label{eq:deepc_reg_constraint_up}
		\hspace{-2.0em}&\mathrm{s.t.}&
                \begin{bmatrix}
			U_{\mathrm{p}} \\ Y_{\mathrm{p}} \\ U_{\mathrm{f}} \\ Y_{\mathrm{f}}
		\end{bmatrix} g =
		\begin{bmatrix}
			u_{\mathrm{ini}} \\ y_{\mathrm{ini}} \\ u \\ y
		\end{bmatrix} +
                \begin{bmatrix}
                        0 \\ \sigma_{y} \\ 0 \\ 0
                \end{bmatrix}, \\
		\label{eq:deepc_reg_constraint_sets}
                \hspace{-2.0em}&~&u \in \mathcal{U}^{N},\ \ y \in \mathcal{Y}^{N}, \\
                \label{eq:deepc_reg_constraint_ini}
                \hspace{-2.0em}&~&u_0 = u_{\mathrm{ini}}(T_{\mathrm{ini}}),\ \ y_0 = y_{\mathrm{ini}}(T_{\mathrm{ini}}).
	\end{eqnarray}
\end{subequations}
Here $r_{t+k}$ is the reference at prediction step $k$, $Q \succeq 0$ penalizes tracking error, and $R \succeq 0$ penalizes input rate changes $\Delta u_k = u_k - u_{k-1}$, where $u_{-1}$ is taken as $u_{\mathrm{ini}}(T_{\mathrm{ini}})$, the last element of the past input window. The constraint sets $\mathcal{U}$ and $\mathcal{Y}$ impose actuator limits and safety bounds over the $N$-step horizon. Constraint \eqref{eq:deepc_reg_constraint_ini} enforces continuity between the past window and the future trajectory: the first predicted values must equal the most recently measured input and output.

\begin{figure}[H]
        \centering
        \includegraphics[width=\columnwidth]{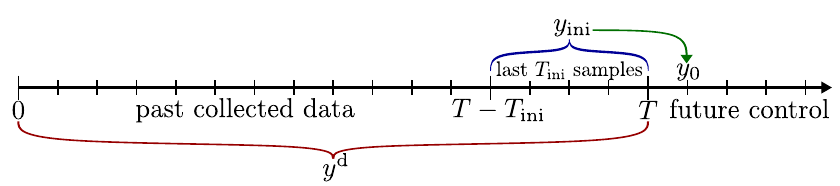}
        \caption{Construction of the DeePC past window of length $T_{\mathrm{ini}}$ from measured data. The extracted suffix defines the initialization variables used in \eqref{eq:deepc_reg_constraint_ini}.}
        \label{fig:deepc_tini_window}
\end{figure}

\begin{figure}[H]
        \centering
        \includegraphics[width=\columnwidth]{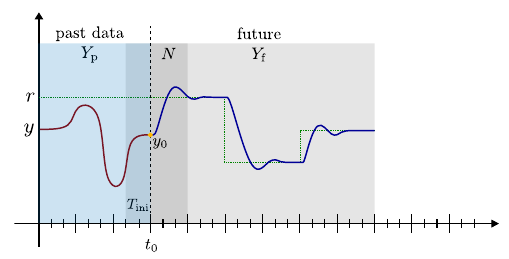}
        \caption{Graphical interpretation of past-output usage in regularized DeePC: the selected past trajectory is matched through \eqref{eq:deepc_reg_constraint_up} (with slack on the output part), while future trajectories are optimized under constraints.}
        \label{fig:deepc_scheme_y}
\end{figure}

Figures~\ref{fig:deepc_tini_window} and~\ref{fig:deepc_scheme_y} show how the past windows $(u_{\mathrm{ini}},y_{\mathrm{ini}})$ are built from measured trajectories and how they enter the behavioral consistency constraint \eqref{eq:deepc_reg_constraint_up}.

The regularization involves two parameters:

\textit{Slack penalty} $\lambda_{\sigma} > 0$: The slack variable $\sigma_y$ relaxes exact matching of past outputs, allowing small deviations when noise or unmodeled dynamics prevent exact satisfaction. In practice, $\lambda_{\sigma}$ is chosen large to preserve fidelity to the data-driven predictor while avoiding infeasibility.

\textit{Ridge regularization} $\lambda_g > 0$: The penalty $\lambda_g\|g\|_2^2$ discourages large coefficient magnitudes, improving numerical conditioning when the Hankel matrix has a high condition number \cite{ref:regularization}. This biases toward coefficient vectors $g$ with small norm, favoring trajectories close to the centroid of the data. Larger $\lambda_g$ increases robustness to noise but can limit adaptation to operating conditions weakly represented in the data.

We solve \eqref{eq:deepc_reg} at each time $t$ in receding-horizon fashion. The initial windows $(u_{\mathrm{ini}}, y_{\mathrm{ini}})$ are taken from the last $T_{\mathrm{ini}}$ measured samples. Algorithm~\ref{alg:deepc} summarizes the procedure.
\begin{algorithm}[H]
\caption{Regularized DeePC}
\label{alg:deepc}
\textbf{Input:} Hankel blocks $U_{\mathrm{p}}, Y_{\mathrm{p}}, U_{\mathrm{f}}, Y_{\mathrm{f}}$, reference $r$, constraint sets $\mathcal{U}$ and $\mathcal{Y}$, weights $Q$, $R$, $\lambda_g$, $\lambda_\sigma$. \\
\textbf{Initialize:} Set $(u_{\mathrm{ini}}, y_{\mathrm{ini}})$ from the last $T_{\mathrm{ini}}$ samples of $(u^{\mathrm d}, y^{\mathrm d})$.
\begin{algorithmic}[1]
\State Solve \eqref{eq:deepc_reg} for $g^\star$.
\State Extract the optimal input sequence $u^\star = U_{\mathrm f}\, g^\star$.
\State Apply $u(t) = u_0^\star$ to the plant and measure $y(t)$.
\State Shift $(u_{\mathrm{ini}}, y_{\mathrm{ini}})$: drop oldest sample, append $(u(t), y(t))$.
\State Set $t \gets t+1$ and \textbf{return to} Step 1.
\end{algorithmic}
\end{algorithm}
\subsection{Koopman-based MPC}

Koopman operator theory provides a framework for representing nonlinear dynamics as linear systems in a lifted (infinite-dimensional) space of observables \cite{ref:koopman_operator}. Under suitable conditions, the nonlinear system can be approximated by a finite-dimensional linear predictor, enabling the use of standard MPC tools for nonlinear systems~\cite{ref:koopman_modeling}. Deep Koopman models~\cite{lusch2018deep} learn the lifting functions via neural networks, yielding Koopman matrices $A_{\mathcal{K}}$, $B_{\mathcal{K}}$, and $C_{\mathcal{K}}$ from data; the identification procedure and architecture are detailed in~\cite{ref:koopman_deepc}. Here we summarize the resulting lifted LTI representation and the offset-free MPC formulation used for comparison. 

The identified deep Koopman model takes the form
\begin{subequations}
\begin{equation}
        z_{k+1} = A_{\mathcal{K}} z_{k} + B_{\mathcal{K}} u_{k},
        \label{eq:kmpc_dynamics}
\end{equation}
\begin{equation}
        y_{k} = C_{\mathcal{K}} z_{k},
        \label{eq:kmpc_output}
\end{equation}
\end{subequations}
where $z_k \in \mathbb{R}^{n_{\mathrm{z}}}$ is the lifted state, $u_k \in \mathbb{R}^{n_{\mathrm{u}}}$ is the input, and $y_k \in \mathbb{R}^{n_{\mathrm{y}}}$ is the output.

To achieve offset-free tracking in the presence of model mismatch and disturbances, we augment \eqref{eq:kmpc_dynamics}--\eqref{eq:kmpc_output} with an output disturbance $d_k \in \mathbb{R}^{n_{\mathrm{y}}}$ assumed constant over the prediction horizon ($d_{k+1} = d_k$), following the standard disturbance-model approach of \cite{ref:offset_free_mpc}. The augmented state $\xi_k = [z_k^\top\; d_k^\top]^\top$ evolves as
\begin{equation}
\label{eq:kmpc_augmented}
\xi_{k+1} = \bar{A}\, \xi_k + \bar{B}\, u_k,\quad y_k = \bar{C}\, \xi_k,
\end{equation}
where
\begin{equation}
\bar{A} = \begin{bmatrix} A_{\mathcal{K}} & 0 \\ 0 & I \end{bmatrix},\;
\bar{B} = \begin{bmatrix} B_{\mathcal{K}} \\ 0 \end{bmatrix},\;
\bar{C} = \begin{bmatrix} C_{\mathcal{K}} & I \end{bmatrix}.
\label{eq:kmpc_aug_matrices}
\end{equation}
A Kalman filter on this augmented system estimates both the lifted state $\hat{z}(t)$ and the disturbance $\hat{d}(t)$ at each time $t$. The KMPC optimization problem is then
\begin{subequations}
	\label{eqn:kmpc_problem}
	\begin{eqnarray}
		\hspace{-2.0em}&\min\limits_{u, z, y}&\sum_{k=0}^{N-1} \left(\|y_{k}-r_{t+k}\|_{Q}^{2} + \|\Delta u_k\|_R^{2}\right), \\
		\hspace{-2.0em}&\mathrm{s.t.}& z_{k+1} = A_{\mathcal{K}} z_k + B_{\mathcal{K}} u_k, \\
		\hspace{-2.0em}&~&y_{k} = C_{\mathcal{K}} z_{k} + d_k, \\
		\hspace{-2.0em}&~&z_0 = \hat{z}(t), \\
                \hspace{-2.0em}&~&d_0 = \hat{d}(t), \\
		\hspace{-2.0em}&~&u_k \in \mathcal{U},\quad y_k \in \mathcal{Y}, \\
		\hspace{-2.0em}&~&\forall\, k \in \{0, 1,\dots, N-1\},
	\end{eqnarray}
\end{subequations}
where $\Delta u_k = u_k - u_{k-1}$ as in \eqref{eq:deepc_reg_objective}, with $u_{-1}$ set to the previously applied input $u(t-T_{\mathrm{s}})$, where $T_{\mathrm{s}}$ is the sampling time of the plant. The weights $Q$, $R$ and constraint sets $\mathcal{U}$, $\mathcal{Y}$ are identical to those in \eqref{eq:deepc_reg}, so that differences in closed-loop behavior are due to the predictive representation, not the tuning. The disturbance estimate $\hat{d}_{t|t}$ is held constant across all prediction steps $k$, which is the standard offset-free MPC formulation \cite{ref:offset_free_mpc}.

\section{Simulation-based Experiments}
\label{sec:sim_experiments}

\subsection{Data Collection}
The dataset for DeePC was generated using a neural-network-based digital twin of the pasteurization unit. To ensure persistent excitation, we applied coordinated random step sequences to all three inputs, with amplitudes selected to cover the full input range. This produced $T = 4000$ samples with sampling time $T_{\mathrm{s}} = 10$ seconds, corresponding to approximately 11 hours of operation. Variable step durations capture both transient and steady-state behavior, enabling identification of full response characteristics. All signals were normalized with standard scalers to improve numerical conditioning.

The theoretical requirements for persistent excitation and data length are introduced in Section~\ref{sec:deepc_willems}. For the selected setup, Hankel construction uses $L = T_{\mathrm{ini}} + N = 70$, giving $T - L + 1 = 3931$ trajectory columns. Persistent-excitation analysis confirmed the rank condition discussed in Section~\ref{sec:deepc_willems} \cite{ref:willems_lemma}. Specifically, $\mathrm{rank}(U_{\mathrm{p}}) = 30$, which is full row rank for $n_{\mathrm{u}}=3$ and $T_{\mathrm{ini}} = 10$.

Note that the DeePC training data is generated from the digital twin, which also serves as the plant in closed-loop experiments. The KMPC model, by contrast, was identified from real experimental data used to build the digital twin (Section~\ref{sec:kmpc_setup}). Both controllers are therefore evaluated against the same simulated plant, but their data sources differ. This means DeePC operates with noise-free trajectories from the model it controls, while KMPC was identified from noisier real measurements. We consider this acceptable because the comparison targets the predictive representation, not the data quality; a fully fair data-pipeline comparison is left to future experimental work.
\subsection{Optimization Problem}
\label{sec:optimization_problem}
The regularized DeePC formulation \eqref{eq:deepc_reg} yields a convex quadratic program (QP) at each control interval. With the Hankel construction from Section~\ref{sec:deepc_willems}, the past blocks are $U_{\mathrm{p}} \in \mathbb{R}^{30 \times 3931}$ and $Y_{\mathrm{p}} \in \mathbb{R}^{30 \times 3931}$, and the future blocks are $U_{\mathrm{f}} \in \mathbb{R}^{180 \times 3931}$ and $Y_{\mathrm{f}} \in \mathbb{R}^{180 \times 3931}$. The resulting QP has 4321 scalar decision variables, 420 equality constraints (past matching and future trajectory definitions), and 720 inequality constraints (box bounds on inputs and outputs over $N$).

Both the DeePC and KMPC optimization problems are solved using the Gurobi solver \cite{ref:gurobi} with warm-starting enabled to exploit temporal continuity between consecutive time steps.

\subsection{Controller Parameters}
\label{sec:controller_parameters}
\begin{table}[H]
\caption{DeePC Controller Parameters}
\label{table:ideepc_params}
\centering
\begin{tabular}{ll}
\toprule
Parameter & Value \\
\midrule
Past horizon ($T_{\mathrm{ini}}$) & 10 \\
Prediction horizon ($N$) & 60 \\
Hankel depth ($L$) & 70 \\
Sampling time ($T_{\mathrm{s}}$) & 10 \\
Output weight ($Q$) & $\mathrm{diag}(20, 0, 0)$ \\
Rate penalty ($R$) & $\mathrm{diag}(20, 20, 20)$ \\
Regularization ($\lambda_g$) & $10^4$ \\
Slack penalty ($\lambda_{\sigma}$) & $10^{5}$ \\
\bottomrule
\end{tabular}
\end{table}

Table~\ref{table:ideepc_params} summarizes the selected controller tuning. The output weight $Q$ prioritizes tracking of $T_1$ only, while $T_2$ and $T_3$ remain constrained. The input rate penalty $R$ is uniform across all three inputs. Parameter definitions are given in Sections~\ref{sec:deepc_willems} and~\ref{sec:deepc_regularized}.

The selected horizons are $T_{\mathrm{ini}} = 10$ and $N = 60$ (600 seconds at $T_{\mathrm{s}} = 10$ seconds), with Hankel depth $L = 70$. This setting balances predictive capability and computational load for real-time optimization with Gurobi \cite{ref:gurobi}.

For comparison, the Koopman-based MPC baseline does not require Hankel-construction parameters ($T_{\mathrm{ini}}$, $L$). It operates on a lifted state-space representation rather than direct input--output trajectories. However, the Koopman controller uses identical weight matrices ($Q$, $R$), prediction horizon ($N$), and sampling time ($T_{\mathrm{s}}$). This ensures a fair comparison between the two data-driven approaches.

\begin{table}[H]
\caption{Input Constraints}
\label{table:constraints}
\centering
\begin{tabular}{lll}
\toprule
Variable & Lower Bound & Upper Bound \\
\midrule
$u_1$ & 30 & 100 \\
$u_2$ & 20 & 100 \\
$u_3$ & 0 & 50 \\
\bottomrule
\end{tabular}
\end{table}

Physical constraints on inputs (Table~\ref{table:constraints}) and outputs are enforced over the prediction horizon. However, the output bounds are naturally satisfied by the reference design and never become active. A barrier-iteration limit of 10,000 and a maximum solve time of 10 seconds are set to support real-time feasibility.

\subsection{Koopman MPC Setup}
\label{sec:kmpc_setup}

The lifted LTI Koopman model~\eqref{eq:kmpc_dynamics}--\eqref{eq:kmpc_output} was identified from the same experimental data used to construct the nonlinear digital twin. The identification yielded matrices $A_{\mathcal{K}} \in \mathbb{R}^{9 \times 9}$, $B_{\mathcal{K}} \in \mathbb{R}^{9 \times 3}$, and $C_{\mathcal{K}} \in \mathbb{R}^{3 \times 9}$, corresponding to $n_{\mathrm{z}} = 9$ lifted states, $n_{\mathrm{u}} = 3$ inputs, and $n_{\mathrm{y}} = 3$ outputs. The lifted dimension was set to $3$ times the output dimension; further increases did not improve performance.

For the offset-free formulation, the output disturbance has dimension $n_{\mathrm{d}} = n_{\mathrm{y}} = 3$, entering only the output equation as described in \eqref{eq:kmpc_augmented}--\eqref{eq:kmpc_aug_matrices}. The augmented system has $\bar{A} \in \mathbb{R}^{12 \times 12}$, $\bar{B} \in \mathbb{R}^{12 \times 3}$, and $\bar{C} \in \mathbb{R}^{3 \times 12}$. A Kalman filter on this augmented model estimates $\xi_k$ with block-diagonal process covariance $Q_{\mathrm{KF}} \in \mathbb{R}^{12 \times 12}$ comprising blocks $0.1\,I_{n_{\mathrm{z}}}$ and $I_{n_{\mathrm{d}}}$, measurement covariance $R_{\mathrm{KF}} = 0.5\,I_{n_{\mathrm{y}}}$, and initial covariance $P_0 = I_{n_{\mathrm{z}}+n_{\mathrm{d}}}$. The filter is initialized with $\hat{z}_0 = C_{\mathcal{K}}^{\dagger} y(0)$ and $\hat{d}(0) = 0$, where $C_{\mathcal{K}}^{\dagger}$ denotes the Moore--Penrose pseudoinverse and $y(0)$ is the measured output at the start of the run. The scalar weights in $Q_{\mathrm{KF}}$ and $R_{\mathrm{KF}}$ were chosen by manual tuning.

The KMPC controller shares the prediction horizon $N$, weights $Q$ and $R$, sampling time $T_{\mathrm{s}}$ (Table~\ref{table:ideepc_params}), and input constraints (Table~\ref{table:constraints}) with DeePC. Both controllers were initialized at the same initial conditions and run for $T_{\mathrm{sim}} = 2000$ time steps with the same reference trajectory.

\subsection{Performance Metrics}
\label{sec:performance_metrics}
Controller performance is evaluated over $T_{\mathrm{sim}} = 2000$ time steps (approximately 5.5 hours). We use the following metrics:

\textbf{Tracking Error:} The root-mean-square tracking error on the controlled output $T_1$:
\begin{equation}
e_{\mathrm{RMS}} = \sqrt{\frac{1}{T_{\mathrm{sim}}} \sum_{t=0}^{T_{\mathrm{sim}}-1} \left(y_1(t) - r_1(t)\right)^2}
\end{equation}

\textbf{Tracking Cost:} The cumulative weighted tracking component of the closed-loop cost:
\begin{equation}
J_{y} = \sum_{t=0}^{T_{\mathrm{sim}}-1} \|y(t) - r(t)\|_Q^2
\end{equation}
Since $Q = \mathrm{diag}(20, 0, 0)$, only $T_1$ contributes.

\textbf{Control Effort Cost:} The cumulative input-rate component, reflecting actuation smoothness:
\begin{equation}
J_{\Delta u} = \sum_{t=0}^{T_{\mathrm{sim}}-1} \|\Delta u(t)\|_{R}^2
\end{equation}
Together, $J_{y}$ and $J_{\Delta u}$ capture the tracking and effort components of the closed-loop cost. For DeePC, the optimization objective also includes the regularization terms $\lambda_g \|g\|_2^2$ and $\lambda_\sigma \|\sigma_y\|_2^2$, which are not included in these metrics.

\textbf{Energy Consumption:} Cumulative electric-heater usage over the simulation:
\begin{equation}
E = \sum_{t=0}^{T_{\mathrm{sim}}-1} u_{3}(t)
\end{equation}
where $u_3(t) \geq 0$ by the input constraints in Table~\ref{table:constraints}. This metric can be linearly transformed to actual energy consumption.

\subsection{Results and Discussion}
\label{sec:results_discussion}

\begin{figure*}[!htbp]
\centering
\includegraphics[width=\textwidth]{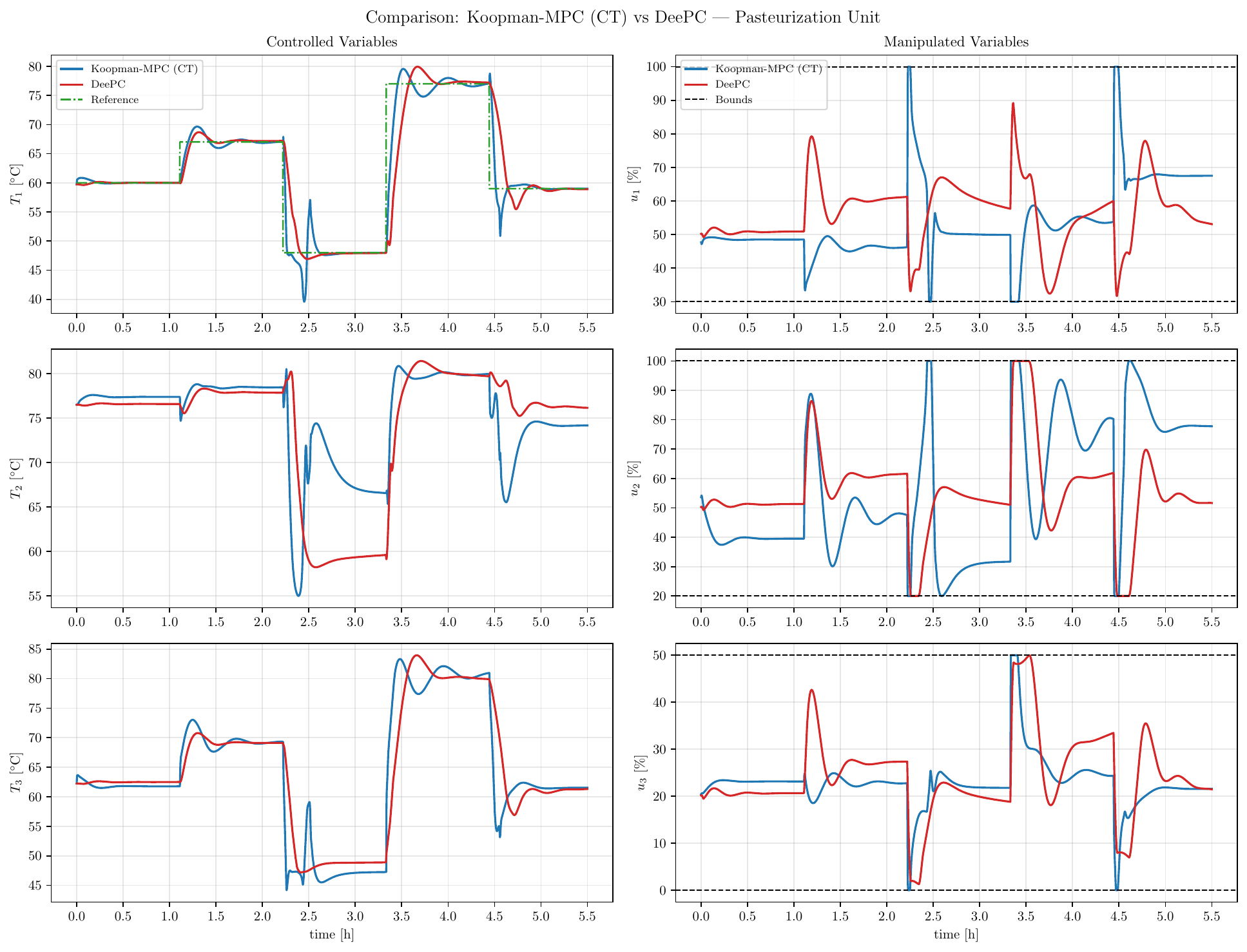}
\caption{Closed-loop comparison of KMPC (blue) and DeePC (red) over approximately 5.5 hours. Left column: outputs $T_1$, $T_2$, and $T_3$. Right column: inputs $u_1$, $u_2$, and $u_3$. The dashed green line is the $T_1$ reference.}
\label{fig:comparison}
\end{figure*}

This section compares closed-loop behavior and evaluates the performance of both data-driven controllers.
Figure~\ref{fig:comparison} provides time-domain behavior while tracking the first output.
Table~\ref{tab:comparison} provides normalized quantitative comparison, where $100\%$ denotes KMPC performance.

\begin{table}[H]
  \centering
  \caption{Comparison of KMPC and DeePC on the pasteurization unit (normalized: $100\%$ corresponds to KMPC performance).}
  \label{tab:comparison}
  \begin{tabular}{lccc}
      \toprule
      Metric & KMPC & DeePC \\
      \midrule
      $e_{\mathrm{RMS}}$    & $100.0$ & $100.7$ \\
      $J_{y}$               & $100.0$ & $223.1$ \\
      $J_{\Delta u}$        & $100.0$ & $16.4$  \\
      $E$                   & $100.0$ & $106.9$ \\
      \bottomrule
  \end{tabular}
\end{table}

From Fig.~\ref{fig:comparison}, both controllers track all reference changes in $T_1$, reaching the target despite following different trajectories. DeePC achieves $e_{\mathrm{RMS}}=100.7\%$ relative to KMPC, confirming comparable tracking accuracy. Although DeePC has no integral action or offset-free mechanism, no systematic steady-state error is observed. This is expected in the current setup: the DeePC training data comes from the same digital twin used for evaluation, so there is no model-plant mismatch or unmeasured disturbance to cause offset. On a real plant, an offset-free extension (e.g., \cite{ref:incremental_control}) would likely be needed.

The strongest contrast appears in control aggressiveness. DeePC generates smoother input trajectories, while KMPC responds faster near reference transitions with sharper actuator moves. This is quantified by $J_{\Delta u}=16.4\%$ for DeePC --- a substantial reduction that is operationally meaningful in terms of actuator wear and process stability.

With three manipulated variables but only a single output weighted in the tracking objective, the closed-loop optimization problem is under-determined with respect to input allocation. As a result, controllers can achieve comparable $T_1$ tracking while producing different input trajectories, as seen in Fig.~\ref{fig:comparison}. 

Under the chosen weight matrix, KMPC achieves better weighted tracking ($J_y = 223.1\%$ for DeePC). The comparable $e_{\mathrm{RMS}}$ but higher $J_y$ for DeePC reflects a different compromise: DeePC sacrifices weighted tracking cost for smoother actuation, rather than performing worse overall.

Energy consumption is similar: DeePC yields $E=106.9\%$, a moderate increase. Despite its much lower control effort cost, DeePC can consume more energy because $E$ measures cumulative heater power (average magnitude), not the rate of input changes. The controller trades energy for smoother actuation.

Overall, both methods are effective for this case study. KMPC is preferable when tight weighted tracking is the priority; DeePC is preferable when smooth actuation matters more. 

   
\section{CONCLUSIONS}

This paper compared two architecturally different data-driven predictive controllers on a multivariable pasteurization unit: DeePC, which eliminates the model step and predicts directly from input--output trajectories, and Koopman-based MPC, which automates model construction via a learned lifted state-space representation. Both controllers shared identical cost weights, horizons, and constraints, isolating the effect of the predictive representation.

The comparison yields a clear, quantitative trade-off. Both methods achieve comparable tracking error ($e_{\mathrm{RMS}}$ within 1\%), but they distribute the cost differently: KMPC delivers $2.2\times$ lower weighted tracking cost $J_y$, while DeePC reduces control effort $J_{\Delta u}$ by $84\%$, producing substantially smoother actuation at a moderate $7\%$ increase in energy consumption. For practitioners, the choice between the two methods reduces to a priority: tighter setpoint tracking or smoother actuator operation.

The paper also provided a self-contained formulation of regularized DeePC, including Hankel construction, regularization design, and a receding-horizon algorithm, applicable to multivariable systems satisfying standard controllability, observability, and persistent excitation assumptions.

Two limitations should be noted. First, DeePC lacks an offset-free mechanism; the absence of steady-state error in the results is a consequence of the noise-free simulation, not a guarantee. Second, all experiments used a neural-network-based digital twin rather than physical hardware. Future work will address both: deploying the controllers on a real pasteurization unit and extending DeePC with offset-free capabilities for operation under measurement noise and plant-model mismatch.

\section*{ACKNOWLEDGMENT}
This research was supported by the Grant Agency of the Czech Technical
University in Prague,
grant No. SGS25/145/OHK3/3T/13.
B. Daráš and M. Klaučo are supported by the European Union project ROBOPROX (Reg. No. CZ.02.01.01/00/22\_008/0004590). 

P. Valábek gratefully acknowledges the contribution of the Scientific Grant Agency of the Slovak Republic under the grants VEGA 1/0239/24 and is supported by an internal STU grant for teams of young researchers and acknowledges the contribution of the EIT Manufacturing – Slovakia – X Fund.

\end{document}